\documentclass{amia}
\usepackage{graphicx}
\usepackage[labelfont=bf]{caption}
\usepackage[T1]{fontenc}
\usepackage[superscript,nomove]{cite}
\usepackage{color}
\usepackage{amsmath}
\usepackage{verbatim}
\usepackage{hyperref}
\usepackage[title]{appendix}
\usepackage{array}
\usepackage{pdfpages}
\usepackage{float}
\usepackage{subcaption}
\usepackage{notoccite}
\usepackage{amsfonts}
\restylefloat{table}
\begin{document}

\title{Learning to Identify Patients at Risk of Uncontrolled Hypertension Using Electronic Health Records Data}

\author{Ramin Mohammadi$^{1,2}$, Sarthak Jain$^{1}$, Stephen Agboola$^{2,3}$, Ramya Palacholla$^{2,3}$,\\ Sagar Kamarthi$^{1,2}$, Byron C. Wallace$^{1}$ }

\institutes{
    $^1$Northeastern University, Boston, MA, USA; $^2$Partners Connected Health Innovation, Boston, MA, USA $^3$ Harvard Medical School, Boston, MA, USA\\
}

\maketitle

\noindent{\bf Abstract}

\textit{Hypertension is a major risk factor for stroke, cardiovascular disease, and end-stage renal disease, and its prevalence is expected to rise dramatically. 
Effective hypertension management is thus critical. 
A particular priority is decreasing the incidence of \emph{uncontrolled} hypertension.
Early identification of patients at risk for uncontrolled hypertension would allow targeted use of personalized, proactive treatments. 
We develop machine learning models (logistic regression and recurrent neural networks) to stratify patients with respect to the risk of exhibiting uncontrolled hypertension within the coming three-month period.
We trained and tested models using EHR data from 14,407 and 3,009 patients, respectively. 
The best model achieved an AUROC of 0.719, outperforming the simple, competitive baseline of relying prediction based on the last BP measure alone (0.634). 
Perhaps surprisingly, recurrent neural networks did not outperform a simple logistic regression for this task, suggesting that linear models should be included as strong baselines for predictive tasks using EHR. 
}  



\section*{Introduction}


Hypertension is a major risk factor for stroke, cardiovascular disease, and end-stage renal disease. In the United States, hypertension has substantial public health and economic implications: it affects 1 out of 3 adults and costs our health system an estimated \$46 billion each year \cite{nguyen}. 
Already a global scourge, the prevalence of hypertension is expected to rise dramatically \cite{mozaffarian}. 
Successful management of hypertension is thus an important objective in light of the substantial cost burden and high rate of adverse outcomes associated with uncontrolled hypertension, and will only increase in importance in coming years. 
A guideline published by the American College of Cardiology defines uncontrolled blood pressure as systolic $\geq$140 or diastolic $\geq$90 \cite{americanHA}. 
Medication non-adherence, unhealthy lifestyle factors, and failure to up-titrate or add anti-hypertensive medications are significant contributors to uncontrolled hypertension.  

Strategic and innovative solutions are needed to improve hypertension management, especially in primary care settings where most patients with hypertension are managed \cite{nguyen}. 
Furthermore, owing to a shift to value-based health-care model, achieving blood pressure control in hypertensive populations serves as an important quality measure for providers. 
Patients at risk for uncontrolled hypertension stand to benefit from early identification as this can trigger proactive treatment regimens and more aggressive education regarding lifestyle modification strategies, as well as use of supportive technologies such as home-blood-pressure monitoring programs. 
However, some of these interventions are costly and cannot be administered indiscriminately to all patients.
Therefore, stratifying patients based on their individual risk for uncontrolled hypertension could help providers make informed treatment decisions and in turn improve long-term outcomes for hypertensive patients.


Prior work has shown that risk stratification can improve outcomes for high risk patients \cite{ogden,kannel}. 
To improve the clinical impact on patients and cost-effectiveness of anti-hypertensive interventions, special attention has to be paid to managing high-risk patients identified through stratification methods \cite{ogden}. 
Traditionally blood pressure (BP) was used to make treatment decisions for managing hypertensive patients. 
However, recent studies indicate that BP alone is not sufficient for making optimal treatment decisions. 
To make informed clinical decisions, clinicians should assess patient risk in light of individual risk factors in addition to BP measurements \cite{ogden}.

Beyond the clinical benefits of decreasing complications due to uncontrolled hypertension and resultant overall quality of life improvement, providers have other incentives to optimize treatment so as to minimize the prevalence of uncontrolled blood pressure. Specifically, under value-based care models, achieving blood pressure control in hypertensive populations is a quality measure (At-Risk Population Hypertension ACO \#28). 
This metric determines the extent to which Medicare reimburses health-care organizations at the end of a given financial year \cite{delaTorreJI}. 
Accountable Care Organizations (ACOs) face significant financial penalties if more than half of their hypertensive population remain uncontrolled at the end of the financial year \cite{JennyGold}. 
Thus, hospitals and care providers have additional incentives to mitigate uncontrolled hypertension and thus meet benchmark standards. 

In closely related prior work upon which we build, Sun \emph{et al.} \cite{sun} developed and evaluated an ML model that predicts transitions between controlled and uncontrolled hypertension and vice versa. Their task formulation is thus slightly different from ours, as their focus is on identifying transition points rather than general risk stratification (i.e., whether someone is likely to be uncontrolled or not in the near future, given current status and other variables extracted from the EHR). But the motivation is ultimately the same. Our findings here largely support those of this prior work \cite{sun}, and this study thus serves as further evidence, derived from a larger corpus, that ML can be used to aid management of hypertension.



\begin{figure}
\centering
\includegraphics[scale=0.4]{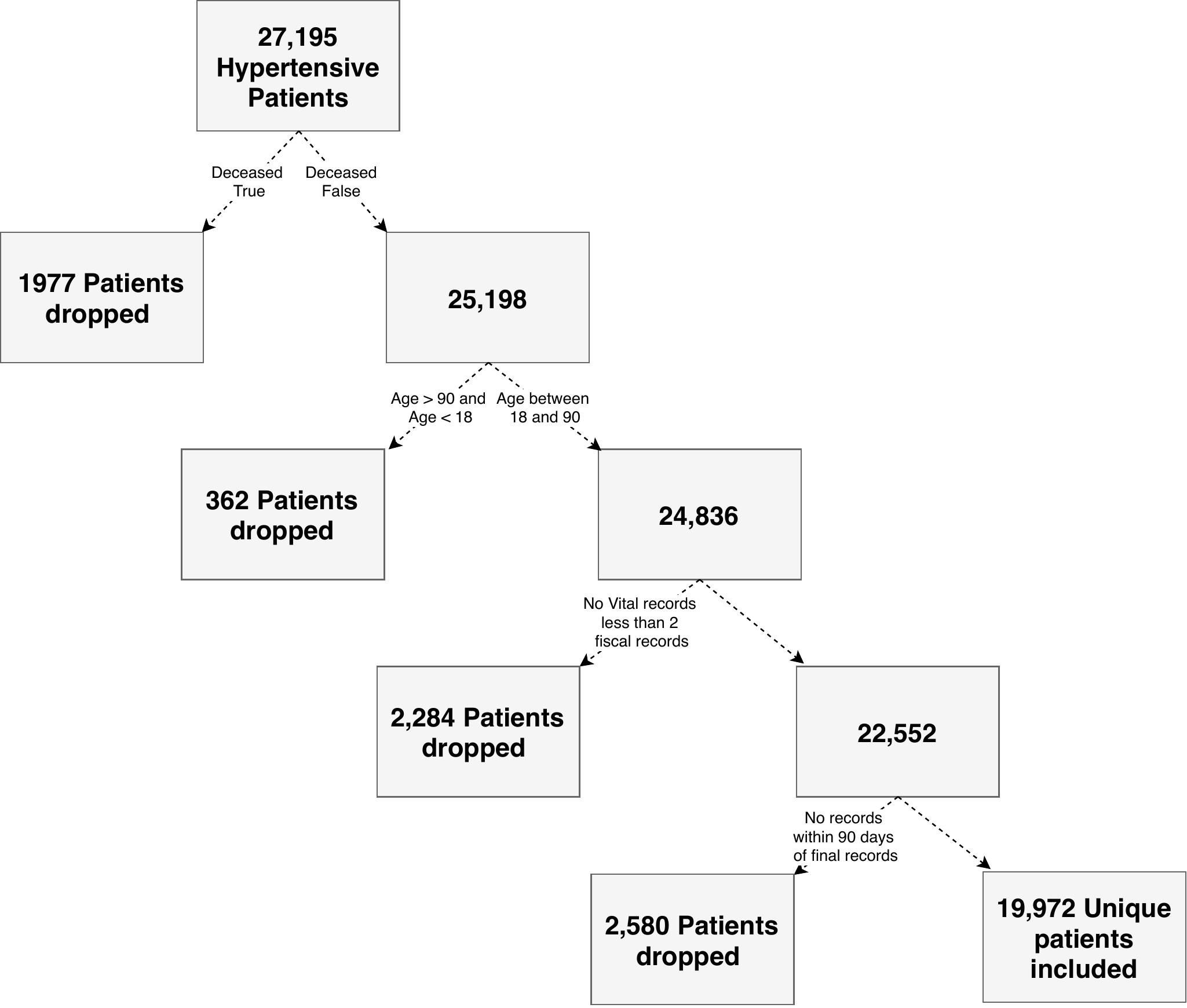}
\caption{The cohort used excludes deceased patients; patients older than 90 and younger than 18; those with fewer than 2 records in a year; and those with no vital sign records.}
\label{figure:Cohort Selection}
\end{figure}

\textbf{Objectives and contributions}. The aim of this work is to empirically evaluate the feasibility of using ML to risk-stratify patients with hypertension with respect to their likelihood of developing uncontrolled hypertension in a fixed time window. Our contributions are as follows. We develop and evaluate models for predicting which patients are likely to fall into the uncontrolled hypertension category within a window of 90 days from their last visit. Using a dataset of EHR from over 27,000 hypertensive patients, we show that ML based approaches outperform the obvious but competitive baseline of simply assuming no change will occur. This evaluation uses a dataset that is an order of magnitude larger that used in prior work \cite{sun}. Also in contrast to this prior effort, we experiment with a modern RNN architecture, namely LSTMs \cite{hochreiter}. However we find that this does not consistently improve performance over a simple logistic regression model.

\section*{Methods}
\label{Methods1}


\textbf{Inclusion Criteria.} 
This is a retrospective analysis of electronic health records (EHR) data. 
We collected data from 27,195 hypertensive patients with approval from the Institutional Research Board (IRB) (protocol number $2016P001661$) at Partners Healthcare. 
Data is from the period of 2010--2016, and includes patients with a primary diagnosis of hypertension. 
We excluded from this pool patients who were deceased, older than 90, or under 18. 
We also excluded patients with fewer than 2 records per fiscal year and/or those with no recorded vital sign data. 
Finally, we excluded patients who did not have any records within 90 days of their last recorded encounter, as this was the predictive window that we deemed operationally feasible. 
Note that this does imply a (potential) bias in our dataset: we are training and evaluating our model on only those patients who had at least two visits within 90 days of each other. 
This resulted in a corpus comprising 19,972 patients in total. 
Figure \ref{figure:Cohort Selection} provides a cohort selection flowchart.

\begin{figure}
\centering
\includegraphics[width=0.8\textwidth]{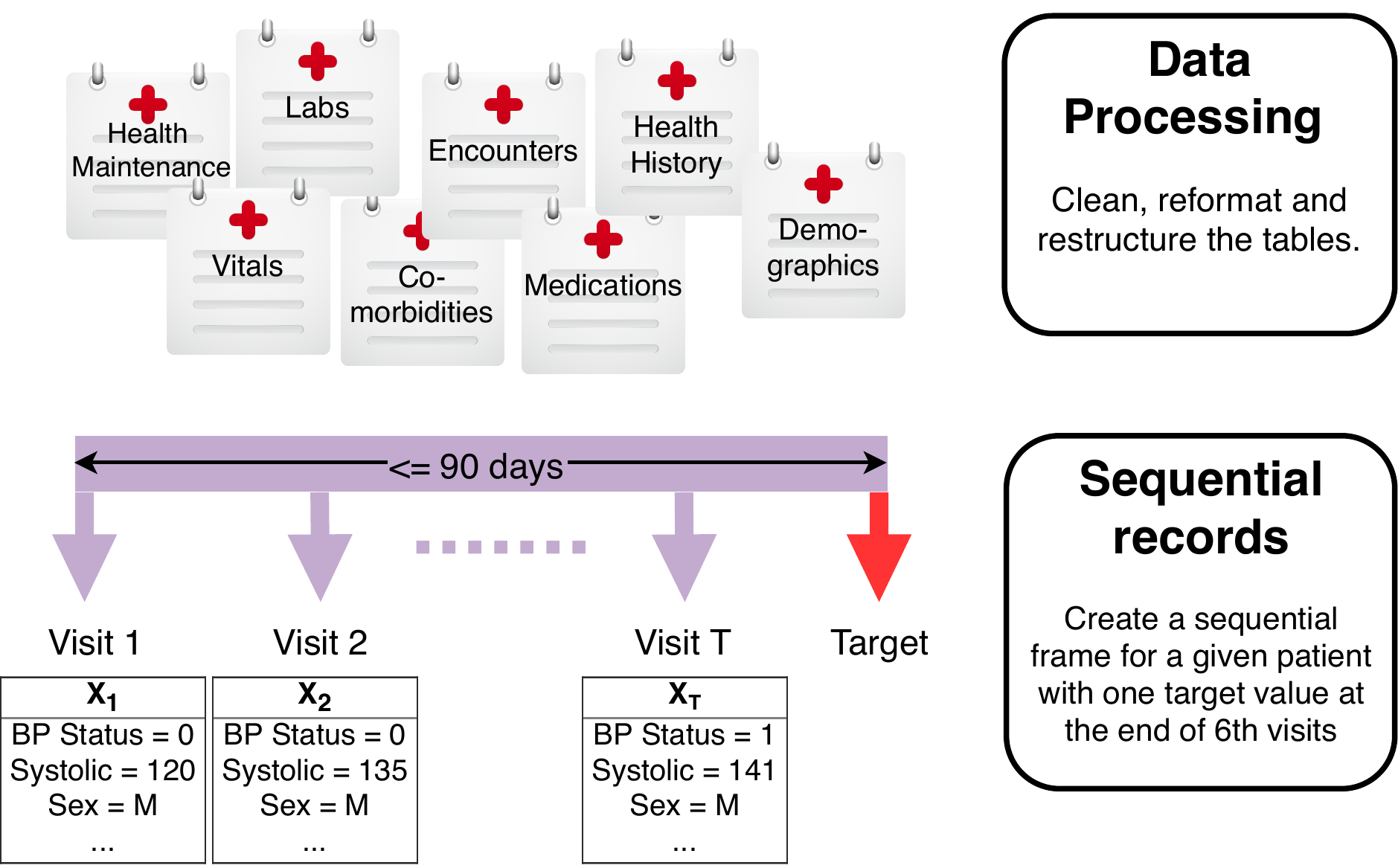}
\caption{A schematic depicting the retrospective predictive task setup we consider. We acquired and cleaned EHR data from all patients in our cohort and created targets that reflect their hypertension status in a ninety window from point of prediction.}
\label{figure:data process}
\end{figure}

\textbf{Design and Feature Engineering} We categorized EHR variables as patient level or hospital level; see Table \ref{table:features}. We grouped patient records into \emph{encounters} which included both inpatient and outpatient visits.


\restylefloat*{table}
\begin{table}
\begin{minipage}{0.48\textwidth}
\centering
\begin{tabular}{ l l } 
\hline
\textbf{Patient Level } &  \\
Demographic information  & Health history          \\
Health information & Vital information       \\
Laboratory test results & Co-morbidities           \\
Medication information  & \\ \hline
\textbf{Hospital Level} &  \\ 
Admission information & Clinician notes       \\ \hline
\end{tabular}
\caption{Features Categories in EHR data}
\label{table:features}
\end{minipage}
\hfill
\begin{minipage}{0.48\textwidth}
\centering
\begin{tabular}{l l l l} 
           & Male & Female & Total\\ \hline
Train      & 6314 & 8093  & 14407\\ 
Validation & 1176 & 1380  & 2556\\ 
Test       & 1742 & 1267  & 3009\\ 
\end{tabular}
\caption{Dataset sample size statistics, in terms of number of patients.}
\label{table:data-distribution}
\end{minipage}
\end{table}

We extracted vitals, medication, health history, problem list and procedure(s) from the encounter records and laboratory orders. Similarly, we used medication codes from medication orders. Encounters are associated with diagnoses lists, encoded as ICD9 and ICD10 codes that we also extracted. For patients with multiple diagnosis codes we considered the principal diagnosis. Medication, problems and lab test orders were coded using binary indicator variables. We report all numerical variables and associated statistics in the Appendix.
We separated the dataset into training, validation, and testing at the patient level, i.e., these sets are disjoint with respect to the patients that they contain. 
We summarize these dataset splits in Table ~\ref{table:data-distribution}.

For numeric variables (e.g., height, pulse) we replaced any missing values with averages taken over patients and/or visits, as appropriate. Records with systolic and diastolic reading less than 90 and 60 respectively were excluded from the study, as these indicate reading errors. 
The blood pressure fraction was defined as systolic over diastolic readings. 
We included lab tests related to hypertension, on the basis of domain expertise. 
We dropped tests with total frequency of less then 60 percent within total records.  


Medications were categorized as: ACE Inhibitor, Diuretic, Beta Blocker, Antihypertensive drug, Calcium Channel Blocker and Vasodilator (Table \ref{tab:Medications}). 
All medications are reported in the Appendix.
Numerical variables were scaled  to range [0, 1], using maximum and minimum values in the training set. 
Variables with more than 99 percent missing values were dropped (see Appendix D and Appendix E). 
Categorical variables were converted to one-hot representation (i.e., indicator vectors).
We labeled patients with systolic BP above 140 or diastolic BP above 90 as \emph{uncontrolled} and others as \emph{controlled}. 
Uncontrolled and controlled statuses were coded as 1 and 0, respectively. 
We fit our model on the training set and chose hyperparameters using the validation set. 
Final model performances were evaluated on the test set, which was completely held-out during development and validation.

The majority of patients (66\%) have controlled hypertension at their target visit. This means our data exhibits \emph{class imbalance}; one target class is substantially more prevalent than another. This can make training discriminative models tricky \cite{wallace2011class}. Here we adjusted class weights associated with targets during training for both models. Specifically, weights for the respective classes were set inverse to their frequencies. 


\textbf{Setting}. We aim to predict which patients will have controlled vs. uncontrolled blood pressure in the near-future, operationally defined here as three months. We cast this as a binary classification task, and evaluated two standard models for such tasks: Logistic Regression (LR) and Long Short Term Memory (LSTM) networks \cite{hochreiter}. 
The latter is a particular type of Recurrent Neural Network (RNN) which has been successfully applied to EHR data in prior work \cite{liptonDiagnose,rajkomar2018scalable}, although to our knowledge not for hypertension specifically.




{\bf Experimental setup}. Prior to any experimentation, we separated the data into training, validation and test sets. These splits were at the patient level, i.e., each patient's records appear in only one of the sets. The test set was used for final evaluation but not used in any way during model development and tuning.

\begin{figure}
\centering
\includegraphics[width=0.7\textwidth]{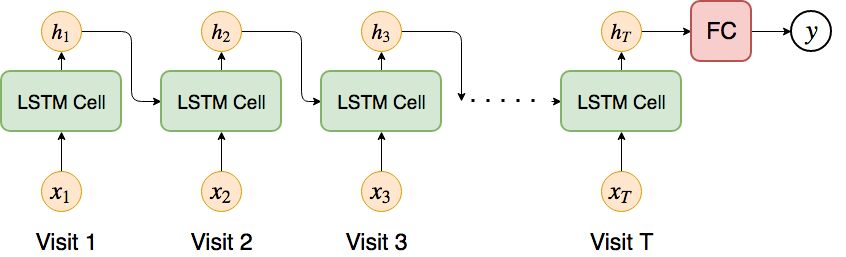}
\caption{LSTM model for processing visits in sequence.}
\label{figure:lstm}
\end{figure}

LSTMs consume sequences of inputs (in our case, a set of ordered vectors encoding information from each visit).
The number of prior visits to pass through the model is a hyperparameter. 
Using the validation set, we found 6 to be the optimal number of records and fed as input sequence to the model. 
We zero-padded sequences corresponding to patients with fewer than 6 encounter records. 
We thus modeled a patient's sequence of records as \(x_1,\cdots,x_T\), where each record $x_i \in \mathbb{R}^F$ is feature vector encoding $F$ features. 
The hidden state obtained from the sequence records is passed to a fully connected layer with sigmoid activation.
Figure ~\ref{figure:lstm} depicts this schematically.

The LR model assumes a single fixed length vector as input from which to make predictions.
Here we use this to encode information extracted from the last patient record, combined with previous blood pressure measurements up to six prior visits. 

For parameter tuning in both models, we performed ad-hoc search over the validation set. 
The $L_1$ regularizer was chosen from range (1e-1,1e-6) and learning rate from range (1e-3, 1e-5). 
For the LSTM model, We first ran the model using Adam optimizer with learning rate (1e-3) then we ran the model for the second time with learning rate of (1e-5). Furthermore, the number of hidden nodes were optimized in range (6, 12, 80 , 120).  
Finally, batch size was chosen from (128,256,512,1024). 

The final optimized LSTM model has one hidden layer with 120 hidden nodes,
dropout rate of 0.2 and 1e-5 penalty for $L_1$ kernel regularizer. 
The optimized LR model uses $L_1$ regularization with a corresponding weight of 0.001 and learning rate of 0.001. 

All models were implemented using Keras \cite{chollet} version 2.2.2 with TensorFlow \cite{abadi} version 1.9.0 and trained on GPU. 
We fit the LSTM using the RMSProp optimizer with binary cross entropy loss. 
For LR, we used the Adam \cite{kingma2014adam} optimizer. 
We used early stopping criteria for assessing convergence, terminating training when loss decreased by $\le 10^{-7}$ on the validation set. 
Under this criterion, the LR model trained for 500 epochs, and LSTM model ran for 250.

\section*{Results} 

We compared developed models against the natural baseline of using the patient's BP measure from their most recent (last) visit as the prediction for current visit. 
This is a reasonably competitive approach because  hypertension status exhibits strong auto-correlation, and our prediction window is relatively narrow (90 days.)

\begin{figure}
\begin{minipage}{0.48\textwidth}
\centering
\begin{tabular}{llll|l}
Model        & Precision & Recall & F1 & AUC  \\ \hline
Baseline & 0.674      & 0.671   & 0.672     & 0.634 \\
LR       & 0.687     & 0.701   & 0.690     & 0.719 \\
LSTM     & 0.696      & 0.713   & 0.700     & 0.714 \\
\end{tabular}
\captionof{table}{Results on the test set.}
\label{table:Test-Results}
\end{minipage}
\begin{minipage}{0.48\textwidth}
\centering
\includegraphics[width=\textwidth]{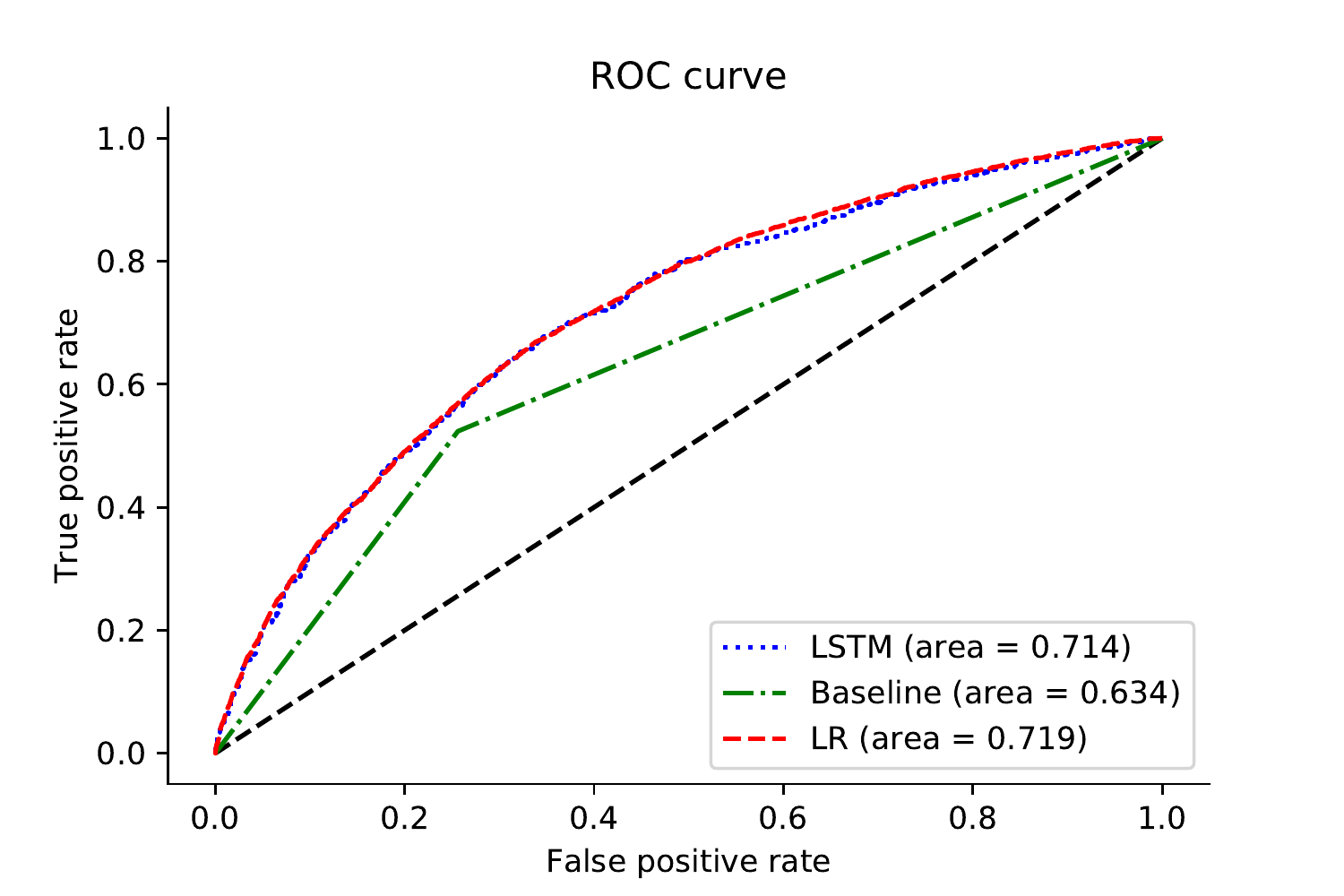}
\caption{ROC curves of each method over the test set. }
\label{figure:Test-Results}
\end{minipage}
\end{figure}

We report results on the test set in Table ~\ref{table:Test-Results}, also summarized in Figure \ref{figure:Test-Results}.

To provide further insights into model predictions we inspect which variables are most responsible for the predictions of a given model. 
In case of LR, a linear model, we simply rank features by (absolute) weight. 
We report the top (highest weighted) 20 variables in Table~\ref{table:LR Top Weights}. 

Inferring the importance of variables in LSTMs is not as straightforward, and multiple options for doing so exist.
Here we adopt a recently proposed method for analyzing deep neural networks, \emph{integrated gradients} (IG) \cite{sundararajan2017axiomatic}. 
This method provides a signed importance score for each variable that reflects its sum contribution to the output.
More concretely, for each data point this method calculates the integral of the gradient of output (i.e., $y$) with respect to each input variable at each time step as we move said variable from the baseline of its current or observed value. 
If the output changes significantly as we vary only one dimension (i.e., the absolute value of integral is large), the corresponding variable is deemed important. 
For additional technical details, we refer the reader to the original paper \cite{sundararajan2017axiomatic}.
We report the top features for the LSTM inferred via IG in Table ~\ref{table:LSTM Top Weights}. We report weights for the top 50 features for both models in the Appendix.


Generally speaking, important features align with intuition. Blood pressure status (controlled vs uncontrolled encoded as 0 and 1 respectively) and systolic BP measurements from prior visits are strongly predictive features in both models, as would be expected.


\begin{table}
\begin{subtable}{0.48\linewidth}
\centering
\begin{tabular}{lr}
{\bfseries Variable Name} &	{\bfseries Weight} \\ \hline
$\text{Systolic}_{(t-1)}$ & 1.492 \\
$\text{Systolic}_{(t-2)}$ & 0.849 \\
$\text{Systolic}_{(t-3)}$ & 0.598 \\
$\text{Blood Pressure}_{(t-1)}$ & 0.550 \\
$\text{Blood Pressure}_{(t-2)}$ & 0.442 \\
$\text{Systolic}_{(t-4)}$ & 0.374 \\
$\text{Blood Pressure}_{(t-3)}$ & 0.349 \\
$\text{Blood Pressure}_{(t-4)}$ & 0.290 \\
$\text{Systolic}_{(t-6)}$ & 0.289 \\
$\text{Blood Pressure}_{(t-5)}$ & 0.268 \\
$\text{Blood Pressure}_{(t-6)}$ & 0.254 \\
$\text{Systolic}_{(t-5)}$ & 0.243 \\
$\text{Blood Pressure}_{(t-7)}$ & 0.226 \\
Mets False & -0.172 \\
BloodLoss False & -0.170 \\
$\text{Systolic}_{(t-7)}$ & 0.152 \\
Smoker & -0.152 \\
Lymphoma False & -0.132 \\
HTN True & -0.130 \\
DSCOP & -0.123 \\ \hline
\end{tabular}
\caption{Top 20 Variables for LR. Subscripts index prior visits.}
\label{table:LR Top Weights}
\end{subtable}
\begin{subtable}{0.48\linewidth}
\centering
\begin{tabular}{lr}
{\bfseries Variable Name} &	{\bfseries Importance} \\ \hline
Time between visits(d) & -0.068 \\
$\text{Systolic}_{(t-1)}$ & 0.024 \\
$\text{Blood Pressure}_{(t-2)}$ & 0.022 \\
$\text{Blood Pressure}_{(t-3)}$ & 0.019 \\
$\text{Blood Pressure}_{(t-1)}$ & 0.019 \\
$\text{Blood Pressure}_{(t-6}$ & 0.018 \\
$\text{Blood Pressure}_{(t-7)}$ & 0.018 \\
$\text{Systolic}_{(t-1)}$ & 0.015 \\
$\text{Systolic}_{(t-7)}$ & 0.013 \\
White & -0.013 \\
Married/Partnered & -0.013 \\
$\text{Systolic}_{(t-3)}$ & 0.012 \\
$\text{Blood Pressure}_{(t-4)}$ & 0.012 \\
Depression False & -0.012 \\
$\text{Systolic}_{(t-4)}$ & 0.012 \\
$\text{Systolic}_{(t-6)}$ & 0.011 \\
Hypertension NOS & -0.010 \\
$\text{Blood Pressure}_{(t-5)}$ & 0.010 \\
BloodLoss False & -0.008 \\
Arrhythmia False & -0.008 \\ \hline
\end{tabular}
\caption{Top 20 Variables for LSTM}
\label{table:LSTM Top Weights}
\end{subtable}
\end{table}

\section*{Conclusion}
All individuals involved in the various aspects of patient care stand to benefit from tools that aid informed clinical decision making. 
From a provider's perspective, identifying which hypertensive patients are likely to become (or remain) uncontrolled can guide targeted, timely interventions and proactive tailored treatments. Thereby, preventing or decreasing the incidence of adverse complications due to uncontrolled hypertension; and improving clinical outcomes and reducing healthcare costs.  

Accurate risk stratification model for hypertension may help increase clinical efficiency, reduce healthcare costs, and improve overall quality of care delivered to hypertensive patients addressing a burgeoning problem in the US healthcare system. 
This work has provided new evidence that ML models can perform this task using a comparatively large dataset of patients, thus complementing existing related work on the problem \cite{sun}. 
We also find, perhaps somewhat surprisingly, that a simple logistic regression model performs about the same as a complex RNN. 
Simple linear models should always be considered as a strong baseline for predictive tasks over EHR. 
 

{\bf Study limitations.} This study has several important limitations, both technical and conceptual.
First, due to the transition of the EHR system to EPIC, there was a gap between medical notes dates and the encounter dates. 
Therefore, we were not able to use notes in the current work; incorporating these may improve the model. 
Second, this retrospective analysis means we had to winnow the set of patients included in the analysis for practical reasons (Figure \ref{figure:Cohort Selection}).  
Third, for simplicity we replaced missing values with simple means, a naive form of imputation. More sophisticated imputation methods, including Bayesian models \cite{sterne2009multiple} and neural network imputation approaches \cite{lipton}, may yield improved performance \cite{sterne2009multiple}. 
We excluded the variables presented in Appendix \ref{Dropped Variables}, due to a high proportion of missing values ($\geq99\%$), which could adversely affect the performance of the model \cite{kotsiantis2007}. 
Few of these excluded variables are likely to be clinically relevant, according to the domain experts involved in this project.
Note that all patients had varying numbers of missing values, but we did not exclude any of them based on missing values (rather, these were simply imputed, as outlined above).

A final potential limitation of this work concerns our creation of target `labels'. 
To do so we required that patients in our cohort had two visits within 90 days (so that the latter of these could serve as the target). 
This excluded 2,580 patients who did not meet this condition. 
This winnowing process may have induced a bias in the sample used for this study, i.e., we cannot be certain that the resultant patient set is representative of the underlying population.

We have here demonstrated that one can achieve reasonably good predictive performance for this task. 
But if such models are to be meaningfully used to inform care, a threshold for clinical action must be established in collaboration with physicians. 

\makeatletter
\renewcommand{\@biblabel}[1]{\hfill #1.}
\makeatother

\bibliographystyle{ieeetr}
\bibliography{main}

\begin{appendices}
\section{Medications}
\begin{table}[H]
\centering
\scalebox{0.7}{%
\begin{tabular}{ l l | l l} 
				\textbf{Drug Family}  &  \textbf{Types} & \textbf{Drug Family}  &  \textbf{Types}\\ \hline
				\textbf{ACE Inhibitor} & Lisinopril, Benazepril &  \textbf{Calcium channel blocker} & Amlodipine, Nifedipine\\ \hline
				\textbf{Diuretic} & Hydrochlorothiazide, Triamterene,Chlorothiazide, & \textbf{Antihypertensive drug} & Nifedipine, Irbesartan,Candesartan,Felodipine, \\
                                  & Hydrochlorothiazide/lisinopril, Chlortalidone  &                                  & Valsartan, Hydrochlorothiazide / Losartan,Telmisartan,\\ 
                                  &                                                &                                  &  Hydrochlorothiazide/lisinopril, Losartan, Chlortalidon \\ \hline
                \textbf{Beta blocker} & Atenolol, Metoprolol, Nadolol, Labetalol, Bisoprolol,  & \textbf{Vasodilator} & Hydralazine \\ 
                                      & Carvedilol                                             & & \\ 
\end{tabular}}
\caption{\label{tab:Medications} Hypertensive Medication}
\end{table}

\section{Results}
\begingroup
\begin{table}[H]
\centering
\scalebox{0.75}{%
\begin{tabular}{lllllllllll}

\multicolumn{11}{c}{\bfseries VALIDATION SET}   \\
\hline

Model & \multicolumn{3}{c}{\bfseries F1-SCORE} & \multicolumn{3}{c}{\bfseries PRECISION} & \multicolumn{3}{c}{\bfseries RECALL} & \multicolumn{1}{c}{\bfseries AUC} \\ 

\cline{2-4} \cline{5-7} \cline{8-10} \cline{11-11} 
        &  {\bfseries Male}  &{\bfseries Female} & {\bfseries Total} & {\bfseries Male}  & {\bfseries Female }& {\bfseries Total}&  {\bfseries Male}  & {\bfseries Female}  & {\bfseries Total}  & {\bfseries Total}                                                                                   \\    \hline
 {\bfseries Baseline}   &  0.52 & 0.76 & 0.68  & 0.51   & 0.77 & 0.68 & 0.52 &  0.76 & 0.68 & 0.68\\       
 {\bfseries LR}   &  0.50&0.80 & 0.70  & 0.58 & 0.76 & 0.70 & 0.43 & 0.84 & 0.70 &  {\bfseries 0.72}\\         
 {\bfseries LSTM}  & 0.47 & 0.81&  {\bfseries0.71}  & 0.55    & 0.77 & 0.70 & 0.41 & 0.85 &   0.72&  {\bfseries 0.72}    \\
 \hline
\multicolumn{11}{c}{\bfseries TEST SET} \\
\hline
 & \multicolumn{3}{c}{\bfseries F1-SCORE} &  \multicolumn{3}{c}{\bfseries PRECISION} & \multicolumn{3}{c}{\bfseries RECALL} &  \multicolumn{1}{c}{\bfseries AUROC} \\

\cline{2-4} \cline{5-7} \cline{8-10} \cline{11-11} 
        &  {\bfseries Male}  &{\bfseries Female} & {\bfseries Total} & {\bfseries Male}  & {\bfseries Female }& {\bfseries Total}&  {\bfseries Male}  & {\bfseries Female}  & {\bfseries Total} & {\bfseries Total}                                                                                        \\    \hline

 {\bfseries Baseline}  &  0.51 & 0.75 & 0.67  & 0.50   & 0.76 & 0.67 & 0.52 &  0.74 & 0.67 & 0.63\\       
 {\bfseries LR}   & 0.49 & 0.79 & 0.69   & 0.57  & 0.75 & 0.69 & 0.43 & 0.84 & 0.70 & {\bfseries 0.72} \\         
 {\bfseries LSTM}   &   0.47 & 0.80 &  {\bfseries 0.70}    & 0.55 & 0.76 & 0.70 & 0.41 & 0.85 & 0.71 &  0.71      \\

\end{tabular}}
\caption {\label{tab:Results} Model Performance per group} 
\end{table}
\endgroup


\section{Variables}
\begingroup
\begin{table}[H]
\centering
\scalebox{0.8}{%
\begin{tabular}{lllll|lllll}

{\bfseries Variable Name} &	{\bfseries Count} &	{\bfseries Mean}	&{\bfseries Std}	& {\bfseries Missing} & {\bfseries Variable Name} &	{\bfseries Count} &	{\bfseries Mean}	&{\bfseries Std}	& {\bfseries Missing} \\
\hline
Heart Rate  & 2223 & 79.36 & 14.69 & 0.98 & Lytes/Renal/Glucose & 21973 & 1.00 & 0.00 & 0.79 \\ 
Height & 98151 & 66.82 & 19.91 & 0.08 & Lytes/Renal/Glucose - POC & 1273 & 1.00 & 0.00 & 0.99 \\ 
Pulse & 93769 & 75.32 & 13.56 & 0.12 & Microscopic Sediment & 2761 & 1.00 & 0.00 & 0.97 \\ 
Respiratory Rate & 22594 & 16.94 & 4.08 & 0.79 & Other Hematology & 2651 & 1.00 & 0.00 & 0.98 \\ 
Temperature & 51541 & 97.92 & 3.20 & 0.51 & Routine Coagulation & 2849 & 1.00 & 0.00 & 0.97 \\ 
Weight & 64057 & 185.58 & 46.09 & 0.40 & Smear Morphology & 1326 & 1.00 & 0.00 & 0.99 \\ 
$\text{Systolic}_{(t-1)}$ & 106125 & 133.30 & 17.26 & 0.00 & Thyroid Studies & 5758 & 1.00 & 0.00 & 0.95 \\ 
$\text{Diastolic}_{(t-1)}$ & 106064 & 75.92 & 10.67 & 0.00 & Tumor Markers & 2114 & 1.00 & 0.00 & 0.98 \\ 
delta time & 106125 & 36.12 & 25.10 & 0.00 & Urinalysis  & 5640 & 1.00 & 0.00 & 0.95 \\ 
BMI & 40008 & 30.20 & 6.43 & 0.62 & Urine Chemistries Random & 3378 & 1.00 & 0.04 & 0.97 \\ 
Fatigue (0-10) & 3100 & 1.67 & 2.88 & 0.97 & $\text{Blood Pressure}_{(t-2)}$ & 106125 & 0.38 & 0.49 & 0.00 \\ 
SexCD & 106125 & 0.43 & 0.50 & 0.00 & $\text{Systolic}_{(t-2)}$ & 106106 & 134.51 & 17.98 & 0.00 \\ 
Age As Of 2010 & 106125 & 61.93 & 13.57 & 0.00 & $\text{Diastolic}_{(t-2)}$ & 106033 & 76.46 & 11.15 & 0.00 \\ 
$\text{BP Fraction}_{(t-1)}$ & 106064 & 1.78 & 0.55 & 0.00 & $\text{Blood Pressure}_{(t-3)}$   & 106125 & 0.35 & 0.48 & 0.00 \\ 
Age Year & 76038 & 65.73 & 13.50 & 0.28 & $\text{Systolic}_{(t-3)}$ & 102869 & 133.90 & 17.60 & 0.03 \\ 
Visit Number & 76038 & 8.32 & 8.94 & 0.28 & $\text{Diastolic}_{(t-3)}$ & 102809 & 76.24 & 10.95 & 0.03 \\ 
Acute Phase Reactants & 1354 & 1.00 & 0.00 & 0.99 & $\text{Blood Pressure}_{(t-4)}$ & 106125 & 0.33 & 0.47 & 0.00 \\ 
Anemia Related Studies & 3188 & 1.00 & 0.00 & 0.97 & $\text{Systolic}_{(t-4)}$ & 99544 & 133.70 & 17.42 & 0.06 \\ 
Blood Diff Absolute & 11573 & 0.79 & 0.41 & 0.89 & $\text{Diastolic}_{(t-4)}$ & 99483 & 76.21 & 10.90 & 0.06 \\ 
Blood Differential \% & 12425 & 0.73 & 0.44 & 0.88 & $\text{Blood Pressure}_{(t-5)}$ & 106125 & 0.32 & 0.47 & 0.00 \\ 
Cardiac Tests & 3241 & 1.00 & 0.00 & 0.97 & $\text{Systolic}_{(t-5)}$ & 96105 & 133.57 & 17.31 & 0.09 \\ 
Complete Blood Count & 14325 & 1.00 & 0.00 & 0.87 & $\text{Diastolic}_{(t-5)}$ & 96051 & 76.14 & 10.85 & 0.10 \\ 
Endocrine Studies & 7581 & 1.00 & 0.00 & 0.93 & $\text{Blood Pressure}_{(t-6)}$ & 106125 & 0.31 & 0.46 & 0.00 \\ 
General Chemistries & 23076 & 1.00 & 0.00 & 0.78 & $\text{Systolic}_{(t-6)}$ & 92725 & 133.55 & 17.29 & 0.13 \\ 
Hepatitis & 1195 & 0.99 & 0.10 & 0.99 & $\text{Diastolic}_{(t-6)}$ & 92670 & 76.13 & 10.85 & 0.13 \\ 
Immunoglobulin & 1179 & 1.00 & 0.00 & 0.99 & $\text{Blood Pressure}_{(t-7)}$ & 106125 & 0.30 & 0.46 & 0.00 \\ 
Lipid Tests & 7049 & 1.00 & 0.00 & 0.93 & $\text{Systolic}_{(t-7)}$ & 89242 & 133.49 & 17.27 & 0.16 \\ 
Liver Function Tests & 12213 & 1.00 & 0.00 & 0.89 & $\text{Diastolic}_{(t-7)}$ & 89188 & 76.10 & 10.88 & 0.16 \\ 

\end{tabular}}
\caption {\label{tab:Variables} Variables statistics} 
\end{table}
\endgroup

\section{Variables Weights}

\begingroup
\begin{table}[H]
\begin{subtable}{0.48\linewidth}
\centering
\scalebox{0.92}{%
\begin{tabular}{lr}
{\bfseries Variable Name} &	{\bfseries Weight} \\ \hline
$\text{Systolic}_{(t-1)}$ & 1.492 \\
$\text{Systolic}_{(t-2)}$ & 0.849 \\
$\text{Systolic}_{(t-3)}$ & 0.598 \\
$\text{Blood Pressure}_{(t-1)}$ & 0.550 \\
$\text{Blood Pressure}_{(t-2)}$ & 0.442 \\
$\text{Systolic}_{(t-4)}$ & 0.374 \\
$\text{Blood Pressure}_{(t-3)}$ & 0.349 \\
$\text{Blood Pressure}_{(t-4)}$ & 0.290 \\
$\text{Systolic}_{(t-6)}$ & 0.289 \\
$\text{Blood Pressure}_{(t-5)}$ & 0.268 \\
$\text{Blood Pressure}_{(t-6)}$ & 0.254 \\
$\text{Systolic}_{(t-5)}$ & 0.243 \\
$\text{Blood Pressure}_{(t-7)}$ & 0.226 \\
Mets False & -0.172 \\
BloodLoss False & -0.170 \\
$\text{Systolic}_{(t-7)}$ & 0.152 \\
Smoker & -0.152 \\
Lymphoma False & -0.132 \\
HTN True & -0.130 \\
DSCOP & -0.123 \\
Drugs False & -0.122 \\
CHF True & -0.105 \\
$\text{Diastolic}_{(t-7)}$& -0.104 \\
Heart Rate missing & -0.103 \\
$\text{Diastolic}_{(t-1)}$ & 0.095 \\
$\text{Blood Pressure}_{(t-7)}$ missing  & 0.095 \\
Height missing & 0.095 \\
$\text{Blood Pressure}_{(t-6)}$ missing & 0.089 \\
Rheumatic False & -0.083 \\
PVD False & -0.083 \\
$\text{Systolix}_{(t-6)}$ missing & 0.082 \\
Blood Differential (\%)  & -0.082 \\
White & -0.076 \\
$\text{Blood Pressure}_{(t-4)}$ missing  & 0.072 \\
PHSOTHER & -0.069 \\
Clinical referral & -0.067 \\
Paralysis False & -0.064 \\
$\text{Systolix}_{(t-3)}$ missing  & 0.064 \\
$\text{Systolix}_{(t-4)}$ missing  & 0.062 \\
PUD False & -0.062 \\
$\text{Systolix}_{(t-7)}$ missing  & 0.061 \\
Anemia False & -0.060 \\
Fatigue (0-10) missing & -0.057 \\
Emergency Flag & -0.054 \\
DDCON & -0.054 \\
FluidsLytes False & -0.052 \\
MARRIED{/}PARTNERED & -0.051 \\
Alcohol False & -0.051 \\
Lisinopril & -0.050 \\
RaceGRP ASIAN & -0.049 \\ \hline
\end{tabular}}
\caption {\label{tab:LR Variables} LR Top 50 Variables Weights} 
\end{subtable}
\begin{subtable}{0.48\linewidth}
\centering
\scalebox{0.92}{%
\begin{tabular}{lr}
{\bfseries Variable Name} &	{\bfseries Weight} \\ \hline
Time between visits(d) & -0.068 \\
$\text{Systolic}_{(t-1)}$ & 0.024 \\
$\text{Blood Pressure}_{(t-2)}$ & 0.022 \\
$\text{Blood Pressure}_{(t-3)}$ & 0.019 \\
$\text{Blood Pressure}_{(t-1)}$ & 0.019 \\
$\text{Blood Pressure}_{(t-6}$ & 0.018 \\
$\text{Blood Pressure}_{(t-7)}$ & 0.018 \\
$\text{Systolic}_{(t-1)}$ & 0.015 \\
$\text{Systolic}_{(t-7)}$ & 0.013 \\
White & -0.013 \\
Married/Partnered & -0.013 \\
$\text{Systolic}_{(t-3)}$ & 0.012 \\
$\text{Blood Pressure}_{(t-4)}$ & 0.012 \\
Depression False & -0.012 \\
$\text{Systolic}_{(t-4)}$ & 0.012 \\
$\text{Systolic}_{(t-6)}$ & 0.011 \\
Hypertension NOS & -0.010 \\
$\text{Blood Pressure}_{(t-5)}$ & 0.010 \\
BloodLoss False & -0.008 \\
Arrhythmia False & -0.008 \\
$\text{Systolic}_{(t-5)}$ & 0.008 \\
AgeAsOf2010 & 0.008 \\
Respiratory rate missing & 0.008 \\
Hypertension NOS & -0.007 \\
Language English & -0.007 \\
Anemia False & -0.007 \\
Rheumatic False & -0.007 \\
Neuro{/}Other False & -0.007 \\
Temperature missing & -0.007 \\
PUD False & -0.007 \\
Weight Loss False & -0.006 \\
Mets False & -0.006 \\
DDCD Other & -0.006 \\
Liver False & -0.006 \\
Hypothyroid False & -0.006 \\
Atenolol & -0.006 \\
DMcx False & -0.005 \\
PVD True & -0.005 \\
AgeYearNBR & 0.005 \\
BMI missing & 0.005 \\
HIV False & 0.005 \\
Pulmonary False & -0.005 \\
Paralysis False & -0.005 \\
$\text{Diastolic}_{(t-1)}$ & 0.005 \\
Blood Diff  Absolute missing & -0.005 \\
Obesity False & -0.005 \\
Coagulopathy False & -0.005 \\
Blood Differential & 0.005 \\
SexCD & -0.004 \\
Abdomnl pain & -0.004 \\ \hline
\end{tabular}}
\caption {\label{tab:LSTM Variables} LSTM Top 50 Variables Weights} 
\end{subtable}
\end{table}
\endgroup

\section{Dropped Variables}
\label{Dropped Variables}
\begin{table}[H]
    \centering
    \begin{tabular}{p{\linewidth}}
         AAA (Abdominal Aortic Aneurysm ) Screening  , ANA Screen  , Albumin/creatinine ratio  , Alcohol Drinks Per Week  , Alcohol Oz Per Week  , Alcohol Use Screening  , Amino Acids  , Amino Acids, urine  , Antibody Screen  , Antiphospholipid Antibodies  , Antiphospholipid Antibody  , Auto-Antibodies  , B12 injection  , Blood Gases/Oximetry  , Blood Pressure-LFA1162  , Blood Type  , Body Surface Area (BSA)  , Bone Marrow Stain  , Bone density  , Breast Exam  , Breast Exam - LHA3537  , Breast Exam - LHA4003  , Breast Exam Instruction  , CRYOs  , CSF Chemistries  , CSF Counts and Diff  , CSF/Fluid, Other  , Calcium Requirements Recommendation  , Carnitine, serum  , Carnitine, urine  , Chlamydia  , Cholesterol  , Cholesterol-HDL  , Cholesterol-LDL  , Cigarettes  , Coagulation Factor Studies  , Colonoscopy  , Complement  , Complete Physical Exam  , Condoms  , Creatinine  , Cystic Fibrosis Carrier  , DNA Diagnostic Tests  , DPT  , DS Glucose  , Dental Exams  , Depo-provera Shot  , Diet  , Diphtheria and Tetanus booster (DT booster)  , Domestic Violence Screening  , Drug Use Screening  , Drugs A-E  , Drugs F-N  , Drugs O-Z  , EGD (upper GI endoscopy)  , EKG  , Echocardiogram  , Exercise Advice  , FEV1-pre (Pre-Forced Expiratory Volume)  , FVC-pre (Pre-Forced Vital Capacity)  , Fetal Activity  , Fluid Chemistries  , Fluid Counts and Diff  , Folic Acid Recommendation  , Foot exam  , Functional Status Screen  , GFR (estimated)  , Glucose  , Gonorrhea  , HCG (Human Chorionic Gonadotropin)  , HCV Ab-LHA3507  , $HIV_x$  , Haemophilus Influenzae type B (HIB)  , Hand Gun Counseling  , HbA1c (Hemoglobin A1c)  , Hct (Hematocrit)  , Head Circumference  , Hearing  , Hemocult x 3  , Hemoglobin Electrophoresis  , Hepatitis A vaccine (Hep A vac)  , Hepatitis B vaccine (Hep B vac)  , Hgb (Hemoglobin)  , HgbAIC  , Home Hemocult  , Home glucose monitoring  , Hypercoagulation Studies  , Hypoglycemia Assessment/Counseling  , INR Result  , Immune globulin  , Inhibitors  , Japanese encephalitis  , KPS (Karnofsky performance status)  , Liver - AST  , Liver - Alkaline Phosphatase  , Liver - Total Bilirubin  , Liver ALT  , Lyme  , Lyme vaccine  , Lymph - \% Difference  , Lymph - Left Arm Volume  , Lymph - Right Arm Volume  , Mammogram  , Measles, Mumps, Rubella (MMR)  , Medicare Annual Wellness Visit  , Meningococcal vaccine  , Microalbumin  , Nutrition Referral  , O2 Saturation - LFA15000  , O2 Saturation - LFA15000.1  , O2 Saturation - LFA12575  , O2 Saturation - LFA38131  , O2 Saturation - LFA38132  , O2 Saturation - LFA4826  , O2 Saturation - LFA4828  , O2 Saturation - LFA5392  , O2 Saturation - SPO2  , OPV / IPV  , On Oxygen?  , Ophthalmology Exam  , Organic Acids, urine  , PSA  , Pain 0-10  , Pain Assessment  , Pain Scale (0-10)  , Pain Score  , Pap Smear  , Peak Flow  , Peak Flow - LHA4483  , Pelvic Exam  , Personal Best Peak Flow  , Platelet Aggregation  , Platelet Antibodies  , Pneumovax  , Podiatry exam  , Positive Antibody Screen  , Pregnancy Weight  , Prepregnancy Height  , Prepregnancy Weight  , Principal ICD Procedure CD  , Prostate exam  , Rabies  , Rabies immune globulin  , Rapid Strep  , Rectal Exam  , Rh Factor  , Routine Serology  , Safe Sexual Practice Counseling  , Seat belt counseling  , Second hand smoke exposure  , Sigmoidoscopy  , Smoking Quit Date  , Smoking Start Date  , Special Coagulation Interp  , Stool Guaiac - 3  , Stool Guaiac-LHA4072  , T-cell Subsets  , TSH-LHA18009  , Testicular Exam  , Testicular Exam Instruction  , Tetanus, Diphtheria,  accellular Pertussis vaccine  , Tobacco Pack Per Day  , Tobacco Used Years  , Toxicology  , Triglycerides  , Trisomy 21  , Tuberculin purified protein derivative  , Typhoid  , UA-Protein  , Urine Chemistries  , Urine Chemistries Timed  , Urine Chemistries Unspec  , Urine Culture  , Urine Dip-LHA4935  , Urine Glucose  , Urine Protein  , Urine Toxicology  , VAS score  , Varicella  , Vision  , Vision-Left Eye  , Vision-Right Eye  , Vitamin D (25 OH)  , Weight Management  , Yellow fever .
    \end{tabular}
    \caption{Variables dropped from consideration due to high proportion of missing values (> 99\%)}
    \label{tab:my_label}
\end{table}

\end{appendices}

\end{document}